\documentstyle{article}
\font\Large=cmr10 at 18truept

\def\Lo{{\cal L}_o}
\def\lg{\log_{10}}
\def\1s{1\sigma}
\def\2s{2\sigma}
\def\3s{3\sigma}
\def\cmsqi{cm^{-2}}
\def\si{s^{-1}}
\def\cmcui{cm^{-3}}
\def\ergsi{erg s^{-1}}
\def\ch2r{\chi_{red}^2}
\begin{document}

\centerline{\bf \Large  Are Gamma-Ray Bursts Cosmological?}
\bigskip 
\bigskip 
\centerline{I. Horv\'ath}
\bigskip 
\centerline{Department of Earth Science, Pusan National University, Pusan,
609-735, Korea}
\centerline{e-mail: hoi@astrophys.es.pusan.ac.kr}

\bigskip 
\bigskip 
\centerline{\bf ABSTRACT} 
\bigskip 

Gamma-ray burst sources are distributed with a high level of isotropy,
which is compatible with either a 
cosmological origin or an extended galactic halo origin. The brightness 
distribution is another indicator used to characterize the spatial 
distribution in distance. In this paper we discuss detailed fits of the 
BATSE gamma-ray 
burst peak-flux distributions with Friedmann models taking into account 
possible density evolution and standard candle luminosity functions. 
A chi-square analysis is used to estimate the goodness of the fits and we
derive the significance level of limits on the density evolution and
luminosity function parameters. Cosmological models provide a good
fit over a range of parameter space which is physically reasonable.

\bigskip 
\bigskip 
\centerline{\bf 1. INTRODUCTION}
\bigskip 

The gamma-ray bursts are distributed isotropically on the sky 
(Fishman et al. 1994; Tegmark et al. 1996; Briggs et al. 1996; although a 
recent paper shows that at least one quadrupole term is non zero, Bal\'azs, 
M\'esz\'aros and Horv\'ath 1998), which is compatible with either a 
cosmological origin or an extended galactic halo origin; 
where the first origin seem to be more probable
(Tegmark et al. 1996, Briggs et al. 1996).
Recently, the successful identifications made
by the Beppo-SAX satellite, followed by the detection of optical counterparts
(van Paradijs et al. 1997),
seem to give a firm support for the models aiming to explain the bursts by
merging neutron stars (Usov and Chibisov 1975; Rees and  M\'esz\'aros 1994;
M\'esz\'aros and Rees 1997)
and seem to put them definitely into the cosmological distances.
In addition, the fainter bursts seem to have longer
durations (Norris et al. 1995; M\'esz\'aros et al. 1996). 
Hence, all these seem to put the bursts definitely into the cosmological 
distances. Nevertheless, further proofs of cosmological origin are still 
useful. The search for a further support is the aim of this paper.

The brightness distribution is the indicator used to characterize the 
spatial distribution in distance, and this can also be used to test the 
cosmological hypothesis. This is generally done by investigation of the 
functional behavior of the integral number $N$ of sources with peak photon
flux rates $P$ above a certain value, $N(>P)$.
Comparisons of observed versus expected values in Friedman cosmologies 
have been discussed, e.g., by Mao and Paczynski (1992), Dermer (1992), 
Piran (1992) and Wasserman (1992). Statistical fits to a $\log N - \log P$
distribution have been done by 
Wickramasinghe, et al. (1993), Cohen and Piran (1994), Emslie and 
Horack (1994), M\'esz\'aros and M\'esz\'aros (1995), Loredo and Wasserman 
(1995), Horack, Emslie and Hartmann (1995), Fenimore and Bloom (1995),
Horv\'ath, M\'esz\'aros and M\'esz\'aros (1996), 
M\'esz\'aros and M\'esz\'aros (1996). 
One of the main questions that such fits must address is the size of the 
parameter space region which is compatible with a cosmological distribution, 
and whether such parameters are reasonable. If the acceptable region 
contains physically plausible parameters and is not too restricted, one may 
assume the consistency of the observations with a general type of models; 
if, on the other hand, the acceptable region is very small and/or populated 
mainly by physically implausible parameters, fine-tuning would be required 
to fit the observations, and the case for consistency with those models is 
weaker. Such consistency, and absence of fine-tuning, is a requirement 
expected of any successful model of the GRB distribution, whether 
cosmological or galactic. Here, we shall address only the question of the 
consistency of the number distribution under the hypothesis of a 
cosmological distribution with a standard candle luminosity distribution.

Most cosmological fits have been made with 
relatively specialized models, generally either with non-evolving or 
evolving density standard candle models, or with non-evolving luminosity 
functions. Limits on the luminosity function were investigated in cosmology 
with a pure density evolution by Horack, Emslie and Hartmann (1995) using a 
method of moments.
In Euclidean space, limits have been investigated by Horack, Emslie and 
Meegan (1994), Ulmers and Wijers (1995) and Ulmer, Wijers and Fenimore 
(1995). Most cosmological calculations have used either the 1B or the 2B 
BATSE data base. In the present paper we make detailed chi-squared 
fits of the observed brightness distribution. 
This is done both using the BATSE 2B catalogue of sources 
(Meegan et al. 1996), and combining the BATSE catalogue with information 
published for the PVO counts (Fenimore and Bloom 1995). The significance 
levels of the various cosmological fits is discussed for both the 2B and 
the expanded burst sample.

Note that, as it seems, there are several subclasses
of gamma-ray bursts
(Kouveliotou et al. 1993; Dezalay et al. 1996; Pendleton et al. 1997;
M\'esz\'aros et al. 1998; Bagoly et al. 1998). Nevertheless,
the criteria are still ambiguous.
Therefore, in this article these
subclasses will not be considered. 

\bigskip 
\bigskip 
\centerline{\bf 2. MODELS AND FITS}
\bigskip 

Analytical expressions for the integral burst number counts $N(>P)$ with
peak photon flux rate in excess of $P$ (units of photon $\cmsqi\si$)
were discussed by M\'esz\'aros  and M\'esz\'aros (1995) and
by M\'esz\'aros  and M\'esz\'aros (1996)  for arbitrary 
Friedmann models with zero cosmological constant .
As discussed in M\'esz\'aros  and M\'esz\'aros (1995), effects of 
a non-flat cosmology ($\Omega_o < 1$) are 
small, and to a first approximation can be neglected. Below we assume
$\Omega_o=1$ everywhere. The effect of a pure density evolution is 
approximated through a dependence
$$
n(z) = n_o (1+z)^D ~, \eqno(1)
$$
where $n$ is the physical burst density rate in $\cmcui yr^{-1}$, $n_o$ is 
the density rate at $z=0$ ($D=3$ corresponds therefore to a 
non-evolving, constant comoving density). 
It is assumed that the sources have the same intrinsic luminosity $\Lo$. 
Therefore the photon 
luminosity function in the 50-300 keV range is represented by the form,
$$
\Phi({\cal L})=    n_o \delta ({\cal L} - {\cal L}_o )   \eqno(2)
$$
For the rest of the paper we take  $P$ to be peak photon flux [$\cmsqi\si$], 
$n_o$ is the physical density of bursts per year at $z=0$. 

The integral number distribution of bursts per year with peak flux rate 
above $P$ is given by M\'esz\'aros  and M\'esz\'aros (1995) as
$$
N(>P)={4\pi \over 3} { \Lo^{3/2} n_0 \over (4 \pi P )^{3/2} }~I, \eqno(3)
$$
where  $I$ is a dimensionless
analytical function of $P$ and the model parameters, i.e. the luminosity
function parameters $\Lo$, the density evolution
parameter $D$, and the density $n(0) = n_o$.

For the numerical fits we used the BATSE 2B catalog (Meegan et al. 1996).
The 1024 ms peak fluxes $P$ (photons $\cmsqi\si$) were used, and only events 
with peak count rates divided by threshold count rates $C_{max}/C_{min} >1$ 
were included, where $C_{min}$ is the published count
threshold for each event (Meegan et al. 1996).
 The 2B sample with this criterion consists of
278 entries in the catalog. Applying the efficiency tables published with the
catalog to correct for detector inefficiency near the trigger threshold, the 
nominal number of bursts accumulated by BATSE over a period of two years 
with peak fluxes above $\log P \geq -0.6$ is 369. 
We chose for these bursts a binning equidistant in $\lg P$, with
step size 0.2 between -0.6 and 1.2, which gives 9 equal bins with a
minimum number of 7 events per bin (in the highest $P$ bin, $\lg P=$
1.0 to 1.2) for the two year 2B sample. The fits were made to the 
differential burst number distribution $N(P)$ as a function of peak photon 
flux $P$ (since only in the differential distribution may the bins be 
considered independent of each other for a $\chi^2$ fit) and the errors 
in each bin were taken to be the square root of the number of events in 
that bin. 

Some of the fits were made using an extended 2B plus PVO sample. For the
PVO events, we used the PVO portion of Table 2 of Fenimore and Bloom (1995)
 for $\lg P \geq 1.2$. A number of subtle issues concerning a matching 
between the different PVO and 
BATSE data sets are discussed by Fenimore et al.  
(1993), who indicate that systematic uncertainties of $\pm 10\%$ in the
relative normalization cannot be ruled out. The matching of the level of the 
BATSE and PVO curves was taken directly from Fenimore and Bloom (1995). 
The PVO data was rebinned, 
ignoring PVO bursts below $\lg P=1.2$ so as not to count twice, and its 
level was renormalized so that the matching 2B data had the same level as 
in the original 2B catalog, i.e about 2 years. The errors for the PVO 
sample were also renormalized taking into account the fact that data had 
accumulated over more than ten years in the PVO case, keeping the relative 
errors the same. We used 5 bins in the PVO range, so that the combined 
2B+PVO fits have 9+5=14 bins, reaching up to $\lg P=3.0$. 

The fits (standard candle with density evolution) involve the
fewest parameters: the photon luminosity $\Lo$ (ph/s), the density $n_o$
and the density evolution parameter $D$.
 For the 2B sample between peak fluxes $-0.6 \leq \log P \leq 1.2$, 
the free parameters are $p=3$  the degrees of freedom are $f=6$ and the 
best $\ch2r$ (reduced chi-square or $\chi^2$ divided by degrees of freedom) 
is 0.85 at the innermost mark. The $\1s,~\2s,\3s$ significance contours were
determined using the standard prescription (e.g., Press, et al., 1986,
or Lampton, Margon and Bowyer 1976). The fit (Fig. a) is good over an 
elongated region describing a relation between the luminosity and the 
density evolution. For faster density evolution $n\propto (1+z)^D$ (larger 
$D$) the luminosity must increase because most sources are at larger 
redshifts, while for slower or negative evolution the luminosity must 
decrease, since most sources are at small redshift ($D=3$ is constant 
comoving density). The optimal fit is obtained for $D=3.5$ and $\Lo\sim 
10^{57} \si$. This luminosity is close to the standard candle value deduced, 
e.g. Horack, Emslie and Hartmann (1995) and Fenimore and Bloom (1995).
(corresponding to $\Lo\sim 10^{51}\ergsi$ for typical
photon energies of 0.5 MeV). However the preference for $D=3.5$ was
not, as far as we can tell, encountered in previous fits. The $\ch2r$
around the best fit minimum is 0.85; however, the $\1s$ region around it is 
rather large, even if not very wide, so this preference is not strong.

The fits using the 2B+PVO sample ($p=3$, $f=11$) are shown in Fig. b, 
with a best $\ch2r=0.62$ at the central mark 
enclosed by its $\1s,~\2s,\3s$ contours. We note that this ignores any 
possible systematic errors in matching BATSE and PVO beyond what is done in
Fenimore et al. (1993), and Fenimore and Bloom (1995). If any extra errors 
were present, they could in principle increase the size of the significance 
regions discussed below (e.g. it might add an extra free parameter for the 
relative normalization). However, such errors are
extremely difficult to quantify without going into additional details 
of the instruments, and we follow Fenimore and Bloom (1995) in adopting 
their relative normalization as adequate without further manipulation.
The effect of the rare high flux PVO bursts satisfying a tight $N\propto 
P^{-3/2}$ correlation at $1.2 \le  \lg P \le  3.0$ is to improve the
best fit (lower $\ch2r$) and to place it at a somewhat smaller
luminosity $\Lo\sim 5\times 10^{56}\si$ and closer to comoving constant
density evolution, $D\sim 3$. This is in good agreement with Fenimore
and Bloom's (1995) value of $L_o \sim 5\times 10^{50}\ergsi$. However, as 
seen from Fig. b, the $\1s$ region around this best fit minimum is compatible
with both larger and smaller $\Lo$ and $D$. In contrast to the pure 2B fit, 
however, the joint $\1s$ upper limit for $\Lo$ and $D$ are $\Lo\le  5\times
10^{57}\si,~D\le  4.5$ (or $\3s$ joint upper limits $\Lo\le  5\times 
10^{58}\si, ~D\le  5$). 

\bigskip 
\bigskip 
\centerline{\bf 3. CONCLUSION}
\bigskip 

The fits presented above show that a cosmological interpretation is 
compatible 
with the data under a variety of assumptions. Good fits of the observed 
differential distribution of bursts $N(P)$ as a function of peak photon flux 
$P$ are obtained under a standard candle 
luminosity function assumption. Fits were obtained for a range of 
density evolution indices $D$, defined through a physical density dependence 
$n_o \propto (1+z)^D$ where $D=3$ is equivalent to a non-evolving, constant 
comoving density case.
The cosmological fits obtained are of good quality ($\ch2r \sim  1$)
for a range of plausible model assumptions. 

The present results are in significant agreement with the literature
(Horack, Emslie and Meegan 1994, Horack, Emslie and Hartmann 1995, 
Ulmer, Wijers and Fenimore 1995, Hakkila et al. 1995,
Horv\'ath, M\'esz\'aros and M\'esz\'aros 1996).

\bigskip 

{\it Acknowledgements:} This research was supported through NASA NAG5-2857,
OTKA T014304 and Sz\'echenyi Foundation Fellowship. 
I am grateful to E. Feigelson,  C. Meegan,  A. M\'esz\'aros, 
P. M\'esz\'aros and J. Nousek for useful advices.

\bigskip 
\bigskip 
\centerline{\bf REFERENCES}
\bigskip 

 Bagoly, Z., et al., 1998, ApJ, 498, 342

 Band, D., et al., 1993, Ap.J. 413, 281

 Bal\'azs, L.G., M\'esz\'aros, A., and Horv\'ath, I., 1998, A\&A,  339, 1

 Briggs, M.S., et al., 1996, ApJ, 459, 40

 Cohen, E., and Piran, T., 1995, Ap.J.(Letters), 444, L25

 Dermer, C.D., 1992, Phys. Rev. Letters, 68, 1799

 Dezalay, J.P., et al., 1996, ApJ, 471, L27

 Emslie, A.G., and Horack, J.M., 1994, Ap.J. 435, 16

 Fenimore, E.E.,, et al., 1993, Nature, 366, 40

 Fenimore, E.E., and Bloom, J.S., 1995, Ap.J. 453, 25

 Fishman, G.J. et al., 1994, Ap.J. (Suppl.), 92, 229

 Hakkila, J., et al., 1995, Proc. ESLAB Symp.

 Horack, J.M., Emslie A.G., and Meegan C.A., 1994, Ap.J. (Letters), 426, L5

 Horack, J.M., Emslie A.G., and Hartmann, D., 1995, Ap.J. 447, 474

 Horv\'ath, I., M\'esz\'aros, P., and M\'esz\'aros, A., 1996, ApJ, 470, 56

 Kouveliotou, C., et al., 1993, ApJ, 413, L101

 Lampton, M., Margon, B., and Bowyer, S., 1976, Ap.J. 208, 177

 Loredo, T. and Wasserman, I., 1995, Ap.J. (Suppl.), 96, 261

 Mao S., and Paczynski, B., 1992, Ap.J. (Letters), 388, L45

 Meegan, C.A., et al., 1996, Ap.J. (Suppl.), 106, 65

 M\'esz\'aros, A., and M\'esz\'aros, P., 1996, Ap.J.  466, 29

M\'esz\'aros, A., Bagoly, Z., Horv\'ath, I., and
M\'esz\'aros, P., 1996, J. Korean Astron. Soc., 29, S43

 M\'esz\'aros, A. et al., 1998. IAU Symposium no. 188. {\bf Hot Universe}, 
 XXIIIrd IAU General Assembly. Kyoto, Japan. Dordrecht: Kluwer Academic, 1998., p.461

 M\'esz\'aros, P. and M\'esz\'aros, A., 1995, Ap.J. 449, 9

 M\'esz\'aros, P., and Rees, M.J., 1997, Ap.J. 482, L29

 Norris, J.P., et al., 1995, ApJ, 439, 542

 Pendleton, C.N., et al., 1997, ApJ, 489, 175

 Piran, T., 1992, Ap.J. (Letters), 389, L45

 Press, W.H., et al., 1986, Numerical Recipes (C.U.P, Cambridge), p.536

 Rees, M.J.,  and M\'esz\'aros, P., 1994, Ap.J. 430, L93

 Tegmark, M., et al., 1996, ApJ, 466, 757

 Ulmer, A., Wijers, R.A.M.J., and Fenimore, E.E., 1995, Ap.J. 440, L9

 Ulmer, A., and Wijers, R.A.M.J., 1995, Ap.J. 439, 303

 Usov, V., and Chibisov, G., 1975, Soviet Astronomy, 19, 115

 van Paradijs, J., et al., 1997, Nature, 386, 686 

 Wasserman I., 1992, Ap.J. 394, 565

 Wickramasinghe W.A.D.T., et al., 1993, Ap.J. (Letters), 411, 55

\vskip 2.0 truecm

\centerline{\bf Figure Caption}
\bigskip 

Cosmological fits for $\Omega_o=1,~\Lambda=0$ for standard candle photon 
luminosity $\Lo$ (photon/s) and physical density evolution 
$n(z)\propto (1+z)^D$.
The inner dots are the best fit $\ch2r$ minimum location, with 
$\1s,\2s,\3s$ contours increasing outwards. 
a) Top: using the BATSE 2B data base; b) Bottom: using the 2B plus 
PVO information.

\end{document}